\documentclass[twocolumn,aps,prb,groupedaddress,letterpaper,amsmath,amssymb]{revtex4}
\pdfoutput=1
\usepackage{graphicx}

\begin{document}

\title{Scanning optical homodyne detection of high-frequency picoscale resonances
in cantilever and tuning fork sensors}

\author{\vspace{-1mm}G. Zeltzer}
\author{J. C. Randel}
\affiliation{Department of Applied Physics, Stanford University,
Stanford, CA 94305}
\author{\vspace{-2mm}A. K. Gupta$^\text{a)}$\footnotetext{$^\text{a)}$Current address: BioMEMS Resource Center, 
Massachusetts General Hospital, Harvard Medical School, Boston, MA 02114}}
\author{R. Bashir}
\affiliation{
Birck Nanotechnology Center, School of Electrical and Computer Engineering, Purdue University, West Lafayette, IN 47907
}
\author{\vspace{-2mm}S.-H. Song$^\text{b)}$\footnotetext{$^\text{b)}$Permanent address: School of Electrical and Electronics Engineering, Chung-Ang University,
Seoul, 156-756, Korea}}
\author{H. C. Manoharan$^\text{c)}$\footnotetext{$^\text{c)}$Electronic mail: manoharan@stanford.edu}}
\affiliation{Department of Physics, Stanford University, Stanford,
CA 94305}

\begin{abstract}
Higher harmonic modes in nanoscale silicon cantilevers and microscale quartz tuning forks are
detected and characterized using a custom scanning optical homodyne interferometer.  Capable of both mass
and force sensing,
these resonators exhibit
high-frequency harmonic motion content with picometer-scale amplitudes
detected in a 2.5 MHz bandwidth, driven by ambient thermal radiation.  Quartz tuning forks
additionally display both in-plane and out-of-plane harmonics.  The first six electronically detected resonances are
matched to optically detected and mapped fork eigenmodes.  Mass sensing experiments utilizing higher tuning fork modes indicate
$> 6\times$\ sensitivity enhancement over fundamental mode operation.
\end{abstract}

\maketitle

Since its invention in 1986 the Atomic Force Microscope (AFM)
has relied on the deflection of a spring-mass system,
originally engineered as a cantilever terminated with a sharp
tip,\cite{Binnig and Quate AFM 1986} responding to very small
external forces exerted between the tip and probed surface.
The introduction in 1991 of the frequency-modulated AFM
scheme allowed enhanced force sensitivity
without a trade-off in operating speed.\cite{Albrecht FM AFM 1991}
Gains in measurement speeds  were achieved by increasing the
fundamental resonant frequency $\omega_0 =
\sqrt{k_\text{eff}/m}$ from the initial $\sim 10^2$ Hz to $\sim 10^5$ Hz.
The sensor spring constants in use have increased from the
originally soft $\sim$ 0.01 N/m for silicon micromachined cantilevers
to much stiffer $\sim$ 3000 N/m quartz tuning fork devices.
The latter proved to be a higher-speed and high-$Q$ alternative,
possessing intrinsic piezoelectric properties characteristic to
quartz crystals.\cite{Giessibl tuning fork 1998}
The piezoelectric response of the fork, linking the mechanical
motion to an electric signal, greatly simplifies AFM design by
allowing a compact and simple readout.
The tuning fork performance is remarkable both at room temperature
as well as in cryogenic environments.\cite{Giessibl atomic resolution 2000}
In most cases the mechanical excited motion of the fork is
the fundamental symmetric in-plane mode.\cite{Rossing92}
In contrast to the fundamental mode,
higher harmonics are predicted to carry information on the
interaction between tip and sample.\cite{Durig 2000}
Electronically detected higher harmonics have led to subatomic
AFM features \cite{Hembacher 2004}
and increased measurement speed.\cite{Liu 2003}
Recently, the use of higher harmonic detection in cantilever
type sensors demonstrated the capability to measure local sample
stiffness\cite{Ozgur Sahin high mode cantilever}
and the contact potential of C$_{60}$ adsorbed on graphite.\cite{Sadewasser06}
Mass detection in cantilever based measurements was shown to
benefit from enhanced sensitivity at higher
modes.\cite{Braun05}
Thus, quantitative knowledge of the higher harmonic content of cantilever and tuning
fork sensors directly feeds into the design of next-generation nanoprobes.

\begin{figure}[h]
\includegraphics[width=1.0 \columnwidth]{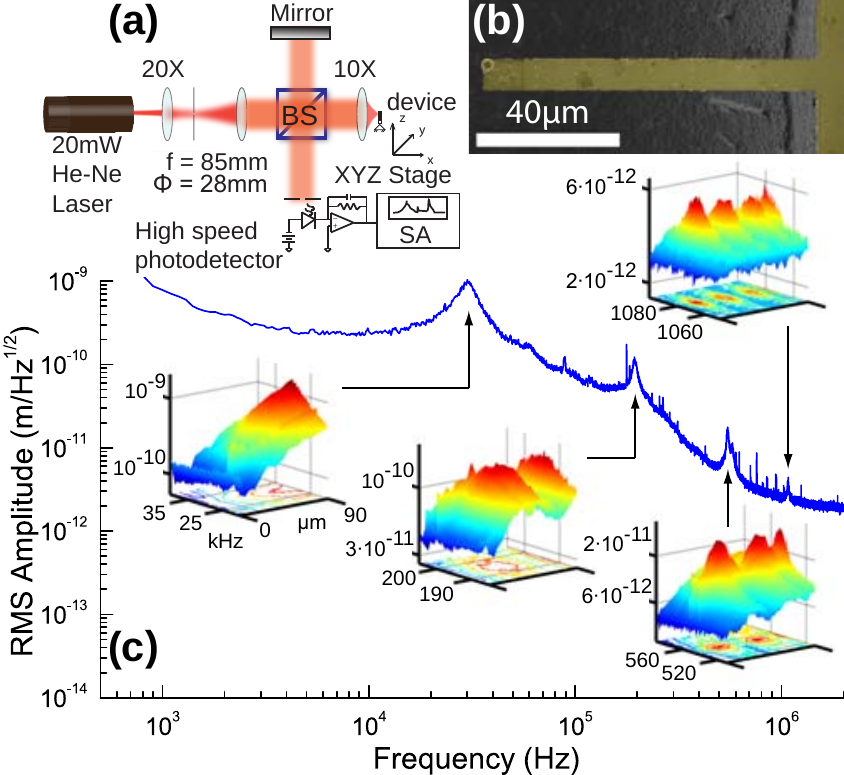}
\vspace{-5mm}
\caption{(Color online) (a) Scanning Michelson interferometer 
including a beam expander (focal length $f$, diameter $\phi$) and focusing lens.
(b) SEM of 200-nm-thick Si cantilever.
(c) High frequency cantilever free-end motion ASD.
(Insets) Cantilever maps of ASD vs position for the first four flexural modes.}
\label{setup}
\end{figure}

In this Letter, we present picoscale characterization based on
a scanning homodyne interferometric scheme allowing detection of high-frequency
oscillation modes in micromachined devices.  Interferometric detection
of cantilever motion has been shown to possess sub-nm sensitivity
in static and low-frequency AFM cantilever deflection\cite{Erlandsson88,Rugar88,Schonenberger89}
as well as in higher bandwidth mass sensing applications.\cite{Lavrik 2003}
Here, a free space Michelson interferometer [Fig.\ 1(a)], with  $\mu$m-scale
spot size in the active arm, enables pm-scale oscillatory motion detection
up to 200 MHz. A coherent and polarized Gaussian beam originating from a 20-mW
HeNe laser ($\lambda = 632.8$ nm) is expanded and collimated
by means of an expansion module delivering a beam waist
$w_\text{beam} = 3$ mm.
The beam is split using a polarizing beam splitter and focused
in the active arm of the interferometer by a $10\times$ microscope objective
($f=15$ mm) to spot size 
$w_\text{spot}=4 \lambda f / 3 \pi  w_\text{beam} = 1.3$ $\mu$m.
A custom built low-noise and high-speed transimpedance amplifier,
with a gain of $10^7$ V/A, amplifies the photocurrent of a reverse-biased
high-speed silicon PIN photodiode allowing low intensity
fringe detection in a 200 MHz bandwidth.
The optomechanical setup is simple and robust to vibrations when
built on a dedicated optical breadboard, offering a relatively
inexpensive solution and measurement accuracy comparable to commercially available
laser Doppler vibrometers.  We apply this technique to sensors of different design
and materials, but matching resonant frequencies.

A benchmark of the measurement capability of such a system is
given in Fig.\ 1(c). Thermally excited modes in a bandwidth of 1.5 MHz
of a silicon cantilever [Fig.\ 1(b)], with length $l=90$ $\mu$m, width $w=8$ $\mu$m, and
thickness $t=200$ nm, are detected as peaks in the amplitude spectral density (ASD) of 
oscillation when the interferometer is positioned at the cantilever free end.
Thermal motion in such devices was previously used to detect femtogram virus
particles binding on the cantilever beam surface.\cite{Gupta}
Using the analytically computed eigenfrequencies of the cantilever
beam,\cite{Sarid}
$f_n = (\alpha_n^2 / 2\pi) \sqrt{Et^2 / 12 \rho l^4}$
with $\alpha_{1\ldots4} = 1.875, 4.694, 7.855, 10.996, \alpha_{5 \dots n}=\pi(n-1/2)$,
mass density $\rho$  and elastic modulus $E_\text{Si(110)}=169$ GPa,
the first four flexural mode frequencies compute to $f_n=
34.0$, 212.8, 596.0, 1167.9 kHz.
These values agree well with measured frequencies
30.6, 194.1, 549.1, and 1080.2 kHz.
Scanning and recording the ASD at 1-$\mu$m  increments along the
cantilever enables mapping of the eigenmodes presented in the insets of Fig.\ 1(c).
An interesting fact observed in the higher modes is the apparent
increase in the $Q$ of each mode from 20 for the fundamental
at 31 kHz to 50 at 1.08 MHz.
Braun et al.\cite{Braun05} have shown that the distributed mass sensitivity of a liquid-submerged cantilever
increases with the mode frequency.  
For our sensors, an enhanced mass sensitivity 
$S=\delta f / \delta m$
is foreseeable at higher modes for
detection of concentrated mass, well beyond $S=0.21$ Hz/fg given by fundamental mode operation only. 
Indeed this sensitivity should further benefit from the fact that the observed $Q$ of the cantilever does not degrade for higher harmonics.

\begin{figure}[h]
\includegraphics[width=1.0 \columnwidth]{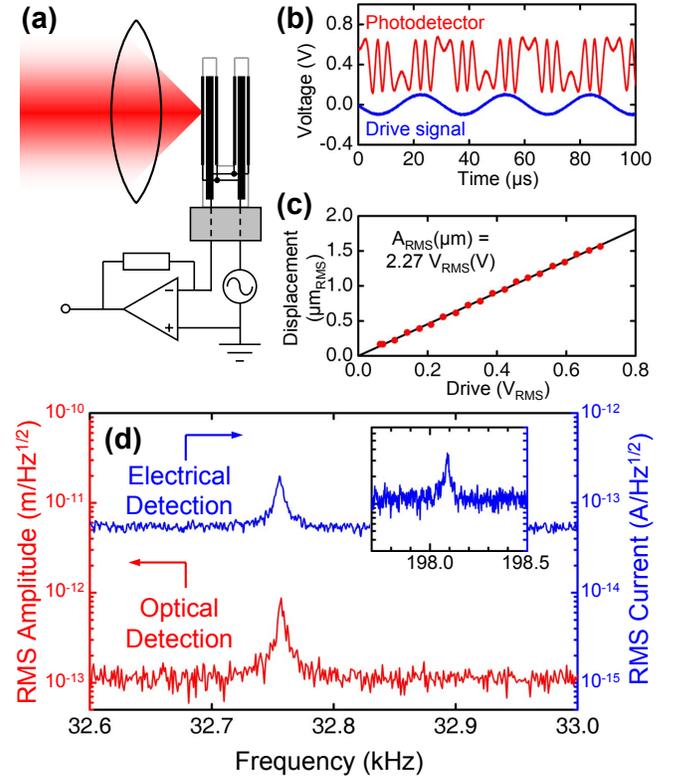}
\vspace{-13mm}
\caption{(Color online) (a) Tuning fork simultaneous optical/electrical measurement setup. (b) Interferometer readout for large fork drive signal.
(c) Tuning fork amplitude vs excitation voltage. (d) Room temperature tuning fork thermal motion detected optically
at the fundamental and electrically at the fundamental and 2$^\mathrm{nd}$ harmonic in-plane modes.\vspace{-3mm}}
\label{tuning fork calibration}
\end{figure}

The tuning fork is an assembly of two cantilevers with intrinsic
piezoelectric response and proven high sensitivity in AFM measurements.
We optically and electrically characterize a
commercial unit with a tine length of $l=2450$ $\mu$m,
width $w=120$ $\mu$m, and thickness $t=220$ $\mu$m.
The theoretical spring constant is  obtained from
$k = E w t^3 / 4 l^3 = 1709$ N/m, where
$E_\text{quartz} = 78.7$ GPa.
The tine displacement is measured by the interferometer
with the active beam arm reflecting off the surface of one of
the tines as shown in Fig.\ 2(a).

Positioning the interferometer at the free end of one
of the fork tines and driving the fork at resonance with
variable excitation voltages allows the calibration of the
fork amplitude response.
In Fig.\ 2(b) the photodetector
signal is recorded while the fork is subjected to
drive signal
$V(t) = V_0 \cos\omega t$ with $V_0 = 90$ mV
and  $f_0 = \omega/2\pi = 32756$ Hz
(the experimentally determined fundamental frequency).


In Fig.\ 2(c) we plot measured amplitude $A$ vs $V_0$ to obtain
the displacement sensitivity $\gamma$ from the slope.  $A$ is related 
to the piezoelectric strain constant  $d_{21}$ and the surface
charge $q$ on each electrode  by
$q/A = 12 d_{21} k l_{e}(l-l_{e}/2)/t^2$
where $l_{e}$ is the electrode length.\cite{Giessibl atomic resolution 2000}
Taking a time derivative and replacing the spring constant $k$ with the
theoretical value, we arrive at an expression for the fork sensitivity
\begin{equation}\label{amplitude vs voltage}
\gamma = \left [ 12\pi d_{21} E f_0 Z
\left ( \frac{wtl_{e}} {l^3} \right)
\left ( l-\frac{l_e}{2} \right)\right]^{-1}
\end{equation}
where $Z$ is the magnitude of the fork's complex impedance.
Setting $d_{21} = 2.31 \times 10^{-12}$ C/N,
$Z=330$ k$\Omega$, $l_{e}=  1.6$ mm, we find $\gamma = 2.8$  $\mu$m/V.
This agrees well with the experimentally observed value of 2.27 $\mu$m/V.
With the fork electrodes shorted, thermally excited motion is
detected optically at the fundamental frequency as shown in Fig.\ 2(d).
The optically detected integrated amplitude is
$x_\text{rms} = 1.58$ pm.  
This is in excellent agreement with the expected value from equipartition,
$x_\text{rms}=\sqrt{k_\text{B} T / k}=1.56$ pm, using $T=300$ K and the theoretical $k$.
Also shown in Fig.\ 2(d) is the simultaneous piezoelectric current
generated by thermally excited motion at
the first two harmonics.

\begin{figure}[h]
\includegraphics[width=1.0 \columnwidth]{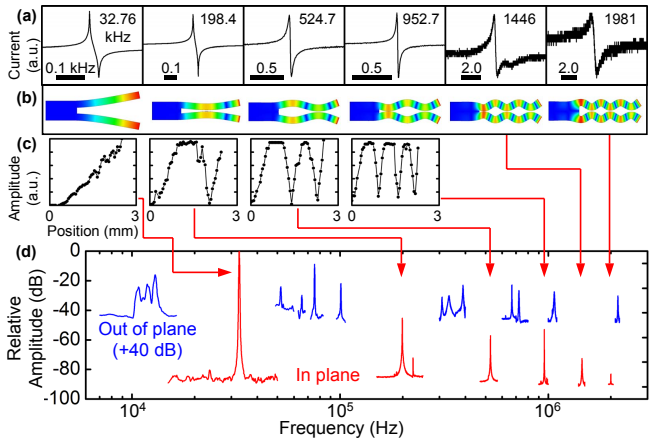}
\vspace{-14mm}
\caption{(Color online) (a) Tuning fork resonances detected via conductance vs drive  frequency  in a 2 MHz bandwidth.
(b) FEA-computed mode shapes for the first 6 symmetric in-plane eigenmodes.
(c) Optically detected mode shapes for the first 4 in-plane eigenmodes.
(d) Fork free-end optically detected in-plane and out-of-plane  resonances using  swept sine excitation.
Amplitudes  normalized to the in-plane fundamental mode. }
\label{spectra}
\end{figure}

We proceed with identifying the higher eigenmodes of the fork,
with both in-plane and out-of-plane amplitude components.
Figure 3(d) displays the ASD of these
detected modes with the readout point at the
free end of the fork.
Treating the fork as a cantilever, the in-plane resonance frequencies are computed
at
32.2, 201.9, 565.3, 1107.7, 1830.9, and 2735.1 kHz, which approximate well
only the first three experimental  values. However, finite-element analysis (FEA) of the entire device
reproduces
the experimental results with better than 5\% accuracy [see computed mode shapes
in Fig.\ 3(b)].  We are able to scan the entire 2.4 mm
tine and map the first 4 modes at 32.756, 198.1, 528.075, and 958.225 kHz
[Fig.\ 3(c)].
The mode shapes for the resonances at 1453.85 and 2000.42 kHz are inferred as the 5$^{\mathrm{th}}$ and 6$^{\mathrm{th}}$
symmetric modes.
Conductance measurements shown in Fig.\ 3(a) were carried out with the fork
placed in vacuum ($\lesssim 0.1$ Torr). The frequencies of these resonances match the first
6 in-plane optical values indicating that the piezoelectric coupling
is maximized at these modes. Electrical excitation of the out-of-plane modes led to
very small amplitudes (in the nm range for driving voltages up to 10 V).
This is equivalent to out-of-plane motion $\sim$ 3 orders of magnitude smaller
than in-plane modes [see Fig.\ 3(d)].

A mass sensing experiment was carried out in vacuum by attaching a 120-ng load to
one tine of the fork.  The first four electrically detected modes experienced
frequency shifts of 50, 148, 221, and 305 Hz, indicating over six-fold 
improvement in mass sensitivity $S$ for operation at the fourth mode over the
fundamental mode.  Compared to Si, the minimum detectable mass with a quartz tuning fork also 
benefits from a $> 100\times$ boost in $Q$ ($> 10^5$ in vacuum).  
An observed $\sqrt{f_n}$ dependence of the $n^\text{th}$ harmonic frequency shift
is currently under investigation.  

We conclude that higher oscillatory modes can be used
to achieve  higher speed and sensitivity for both force and mass sensing
applications.  The experimental mode shape identification can be used in optimizing sensing
tip and mass position within nanoprobe setups.

This work was supported by NSF (Stanford-IBM Center for Probing the Nanoscale and CAREER Program) 
and ONR (YIP/PECASE).
We acknowledge fellowship support from NSF (J.C.R), SBS Foundation (S.-H.S), and the Alfred P. Sloan Foundation (H.C.M).
We thank D. Weld and T. Kopley for discussions.

\bibliographystyle{aip}

\begin{thebibliography}{2}

\bibitem{Binnig and Quate AFM 1986}
G. Binnig and C. F. Quate,
\newblock Phys. Rev. Lett. {\bf 56}, 930 (1986).

\bibitem{Albrecht FM AFM 1991}
T. R. Albrecht, P. Gr\"{u}tter, D.  Horne, and D. Rugar,
\newblock J. Appl. Phys. {\bf 69}, 668 (1991).

\bibitem{Giessibl tuning fork 1998}
F.  J. Giessibl,
\newblock Appl. Phys. Lett. {\bf 73}, 3956 (1998).

\bibitem{Giessibl atomic resolution 2000}
F.  J. Giessibl,
\newblock Appl. Phys. Lett. {\bf 76}, 1470 (2000).

\bibitem{Rossing92}
T. D. Rossing, D. A. Russel, and D. E. Brown,
\newblock Am. J. Phys. {\bf 60}, 620 (1992).

\bibitem{Durig 2000}
U. D\"{u}rig,
\newblock New J. Phys. {\bf 2}, 5.1 (2000).

\bibitem{Hembacher 2004}
S. Hembacher, F. J. Giessibl, and J. Mannhart,
\newblock Science {\bf 305}, 380 (2004).


\bibitem{Liu 2003}
S. Liu, J. L. Sun, H. S. Sun, X. J. Tan, S. Shi, J. H. Guo, and J. Zhao,
\newblock Chin. Phys. Lett. {\bf 20}, 1928 (2003).

\bibitem{Ozgur Sahin high mode cantilever}
O. Sahin, C. F. Quate, and O. Solgaard,
\newblock Phys. Rev. B {\bf 69}, 165416 (2004).

\bibitem{Sadewasser06}
S. Sadewasser, G. Villanueva, and J. A. Plaza,
\newblock Appl. Phys. Lett.  {\bf 89}, 033106 (2006).

\bibitem{Braun05}
T. Braun, V. Barwich, M. K. Ghatkesar, A. H. Bredekamp, C. Gerber, M. Hegner, and H. P. Lang,
\newblock Phys. Rev. E {\bf 72}, 31907 (2005).


\bibitem{Rugar88}
D. Rugar, H. J. Mamin, R. Erlandsson, J. E. Stern, and B. D. Terris,
\newblock Rev. Sci. Inst. {\bf 59}, 2337 (1988).

\bibitem{Schonenberger89}
C. Schonenberger and S. F. Alvarado,
\newblock Rev. Sci. Inst. {\bf 60}, 3131 (1989).

\bibitem{Erlandsson88}
R. Erlandsson, G. M. McClelland, C. M. Mate, and S. Chiang,
\newblock J. of Vac. Sci. Technol. A {\bf 6}, 266 (1988).

\bibitem{Lavrik 2003}
N. V. Lavrik and P. G. Datskos,
\newblock Appl. Phys. Lett. {\bf 82}, 2697 (2003).

\bibitem{Gupta}
A. Gupta, D. Akin, and R. Bashir,
\newblock Appl. Phys. Lett. {\bf 84}, 1976 (2004).

\bibitem{Sarid}
D. Sarid,
\newblock \textit{Scanning Force Microscopy} (Oxford University Press, New
York, 1994).


\end{thebibliography}

\end{document}